\documentclass[runningheads]{llncs}
\usepackage{booktabs} % For formal tables

\usepackage{epsfig,endnotes}
\usepackage{paralist}
\usepackage{graphicx} % for pdf, bitmapped graphics files

\usepackage{url}
\usepackage{verbatim}
\usepackage{tabularx}
\usepackage{subcaption}
\usepackage{multirow,color}

\usepackage[shortlabels]{enumitem}
\setitemize{noitemsep,topsep=0pt,parsep=0pt,partopsep=0pt}
\setenumerate{noitemsep,topsep=0pt,parsep=0pt,partopsep=0pt}

\usepackage[subtle]{savetrees}
\graphicspath{{./figures/}}

\usepackage{amssymb}
\usepackage[cmex10]{amsmath}

\begin{document}
\title{Fighting Voice Spam with a Virtual Assistant Prototype\vspace{-20pt}}
%\title{\Large \bf Fighting Voice Spam with a Virtual Assistant Prototype}

\author{Sharbani Pandit\inst{1} \and
Jienan Liu\inst{2} \and
Roberto Perdisci\inst{1,2}\and
Mustaque Ahamad \inst{1}}
\authorrunning{S. Pandit et al.}
\institute{Georgia Institute of Technology \\
\email{pandit@gatech.edu}\\
\email{mustaq@cc.gatech.edu}
\and
University of Georgia\\
\email{jienan.liu25@uga.edu}\\
\email{perdisci@cs.uga.edu}
}
%\author{

%{\rm Anonymous}
% copy the following lines to add more authors
% \and
% {\rm Name}\\
%Name Institution
%} % end author

\maketitle
\vspace{-20pt}
\begin{abstract}
%\boldmath
Mass robocalls affect millions of people on a daily basis. Unfortunately, most current defenses against robocalls rely on phone blocklists and are ineffective against caller ID spoofing. To enable the detection of spoofed robocalls, we propose a {\em virtual assistant} application that could be integrated on
smartphones to automatically vet incoming calls. Similar to a human assistant, the virtual assistant can pick up an incoming call and screen it without user interruption to determine if the call is unwanted. Via a user study, we show that our virtual assistant is able to preserve the user experience of a typical phone call. At the same time, we show that our system can detect mass robocalls without negatively impacting legitimate callers. 
\end{abstract}

% \keywords{Telephone Spam, Virtual Assistant, Robocall Detection, Usability}

% \fi
%%\IEEEpeerreviewmaketitle

\section{Introduction}
Telephony has been a relatively secure channel for voice communication for over 140 years. Currently, about 4.77 billion people across the world rely on the global telephony system for voice communication. Recent technological advances (e.g., Voice over IP) and policy changes offer many benefits, including lower cost and richer phone call experience. However, due to its convergence with the Internet, the phone channel has become a target of attacks by fraudsters and cyber criminals \cite{c0,c1}. Cheap mass robocalling \cite{c4}, voice phishing \cite{c2} and caller ID spoofing \cite{c3} are some of the techniques that are being used by fraudsters in these attacks. The number of robocalls continues to hit new highs each year. The anti-robocalling company YouMail estimates that February 2020 saw 4.8 billion robocalls~\cite{source} in the United States. Other companies have recorded similar high volumes of robocalls as well~\cite{wired}, and complaints about unwanted calls in August 2019 to the Federal Trade Commission (FTC) in USA  numbered about 428,000~\cite{c12}.

\par In response to the increasing number of unwanted phone calls, a number of call blocking smartphone applications have appeared such as Hiya, Truecaller, YouMail etc. which are used by millions of users. Although such call blocking apps can be somewhat effective against scam calls, their effectiveness can be easily degraded with caller ID spoofing \cite{blacklist-paper}. Such spoofing is easy to achieve, and robocallers have resorted to tricks like neighbor spoofing (caller ID is similar to the targeted phone number) to overcome call blocking and to increase the likelihood that the targeted user will pick up the call. To help reduce caller ID spoofing, both industry groups and regulatory bodies have explored stronger authentication for call sources. Although the recently published IETF RFC 8588 describes SHAKEN/STIR approach to enhance trust in the source of a call with signatures \cite{stir,shaken,strike-force}, its widespread deployment by carriers will likely take many years. Furthermore, elimination of caller ID spoofing will not make all unwanted calls go away, as phone numbers can be cheaply acquired and used to overcome blacklists.

\par At a high level, the robocall problem resembles the email spam problem, in which information about the source of an email could potentially be spoofed. Over the years, the security community has been successful in developing effective spam filtering solutions to ensure that  email remains a viable means of communication \cite{network,snare}.  However, most techniques used in such solutions cannot be applied to detect and filter voice spam. This is because when a user receives a phone call, the only information the user can rely on before answering is the caller ID (i.e., the calling phone number). Namely, the {\em context} of the call (i.e., the content of the caller's message) cannot be used to detect if the call is spam or not.

\par To detect unwanted robocalls and to provide the user with more meaningful call context when a phone rings, compared to only relying on the (spoofable) caller ID, we introduce RobocallGuard, a natural voice interaction model which is mediated by a Virtual Assistant (VA). The VA mimics a human call screener (e.g., a secretary) who picks up an incoming phone call and makes the user aware of the call only when it confirms that the call is not a robocall or other type of spam. When a call arrives, if the caller ID is not among the user's contact list, the VA transparently picks up the call and briefly interacts with the caller to determine if its source is a robocaller. Such interaction aims to be natural for legitimate callers, while enabling the detection of robocall sources who indiscriminately target a large number of victims. Furthermore, such interaction with the VA enables learning the context of the call. Calls that are not detected as spam are passed on to the user, and the context extracted from the conversation between the VA and the caller is provided simultaneously, allowing the user to make an informed decision on whether the call is unwanted or legitimate.

\par Recently, a number of automated caller engagement systems that attempt to collect information about a call source have been proposed. The \textit{Call Screen} feature available on the latest versions of the Android phone app provides call context to the user via a real time transcript of the call audio. When an incoming call arrives, the user is prompted with three options: answer, decline and screen. If the screen option is chosen, Google Assistant engages with the caller to collect audio and generate a transcript of the ongoing call. However, users are notified (i.e., the phone rings) of all incoming calls (including robocalls) and user intervention is needed to screen such calls. Google Voice allows screening calls to hear the name of the caller before users answer the call. In contrast, Robokiller \cite{robokiller}, a smartphone application, features an \textit{Answer Bot} that detects spam calls by forwarding all incoming calls to a server, which accepts each call and analyzes its audio to determine if the audio source is a recording. Once the call is determined to come from a human, it is forwarded back to the user. In Robokiller, a caller continues to hear rings while the call is picked up, analyzed and forwarded back to the user, which could negatively impact legitimate callers. Also, the audio analysis techniques used by Robokiller can be countered by a more sophisticated robocaller, and unwanted calls originating from human callers, such as telemarketers or human callers hired by scam campaigns like Tech support, IRS etc., cannot be  stopped.

To the best of our knowledge, no systematic usability and effectiveness studies have been reported of either Robokiller or Google's Call Screen. Our goal is to explore an automated voice-based interaction approach that maintains both caller and callee user experience, eliminates user interruption and stops unwanted calls even in the presence of spoofed calls. We evaluate RobocallGuard's detection capabilities with a corpus of real robocalls and conduct a user study to evaluate its usability.

\par Although it may not be possible to stop all unwanted calls, we believe more secure communication via the telephony channel can be supported by an automated call screening agent that can detect and block such calls without degrading user experience. The voice interaction model investigated in this paper aims to help achieve this goal and we provide a proof-of-concept demonstration that it could be easily supported by current smartphones.

\par In summary our paper makes the following contributions.
\begin{compactitem}
    \item To the best of our knowledge, we are the first to evaluate a call screening virtual assistant that uses automated call handling and audio analysis to defend against robocalls and other types of spam calls, including those that evade blacklists with caller ID spoofing. In addition, transcription of call audio recorded by the virtual assistant is used to provide meaningful context about incoming calls to a user when the phone rings.
    \item  Our virtual assistant aims to provide a mechanism that is similar, albeit much less ``sophisticated'', to having a human call screener who can pick up phone calls and only forward to the user those calls which are likely wanted and record messages for the calls that are most likely unwanted. As a result, users are not annoyed with continuous ringing from unwanted calls.
    \item To demonstrate the ability of the virtual assistant to detect robocalls, we have developed a proof-of-concept smartphone app named \emph{RobocallGuard}. To this end, we experimented with a corpus of 8,000 real robocalls collected by a large phone honeypot, and show that all of them can be detected and thus blocked.
    \item  In addition, our proof-of-concept app allowed us to conduct an institutional review board (IRB)-approved user study to assess the usability of our virtual assistant. The results of this study demonstrate that the natural experience of a typical phone call is preserved for both callers and receivers, while benefiting from the ability to detect robocalls and other potentially unwanted calls.
\end{compactitem}

\begin{comment}
\par The rest of the paper is structured as follows. Section \ref{sec:related_work} discusses related work and how our approach can be distinguished from other similar work. Section \ref{sec:sys_design} describes the threat model assumed in our work and presents the
system design of our voice interaction model. An implementation of a proof-of-concept prototype of the VA is described in Section \ref{sec:implementation}. Section \ref{sec:results} presents usability study  and performance evaluation of RobocallGuard. We discuss the limitations and answer
some questions that readers may have about our system in Section \ref{sec:discussion}. Finally, we explore challenges associated with detection of more sophisticated attacks that will come over the voice channel in the future and conclude our work in Section \ref{sec:conclusion}.
\vspace{-1ex}
\end{comment}

\section{Related Work} \label{sec:related_work}

\par Abuse in the telephony channel has grown considerably and high volume scams
that rely on this channel have proliferated in the recent years \cite{c8,c11,c9,tech-support,bharat,irs}. 

\begin{comment}
Security researchers have explored such abuse and specific scams. For example, Costin et. al. \cite{c9} investigated how phone numbers are used in various scams based on the analysis of crowd sourced web listings. The tech support scam has been one of the most prominent and have received attention from regulators and law enforcement. Insights into the online and phone infrastructure used by tech support scams and their tactics have been explored in \cite{tech-support,bharat}. However, the work on understanding scams only considered phone numbers related to outgoing calls. The targeting of international students by the IRS scam in the United States was explored in \cite{}. 
\end{comment}

\begin{comment}
Robocalling is widely recognized as a serious problem and has received
significant attention from industry and regulatory bodies. For example,
commercially available apps such as Truecaller \cite{truecaller}, Nomorobo
\cite{nomorobo}, Youmail \cite{youmail}, Hiya \cite{hiya} etc. provide defenses
against robocalls. Telephone carriers such as AT\&T \cite{at&t}, Verizon
\cite{verizon} etc, are also providing users with ``call protect" options which
can screen incoming phone calls. These commercially available solutions
generally use metadata such as source phone number and use blacklists of known
fraudulent robocallers to block unwanted calls and alert users about them.
Blacklists rely on historical data such as user complaints or honeypot generated
information \cite{blacklist-paper, bharat-paper, c10}. 
\end{comment}

It has been shown that phone blacklisting methods provided by smartphone apps
(e.g., Truecaller \cite{truecaller}, Nomorobo \cite{nomorobo}, Youmail
\cite{youmail}, Hiya \cite{hiya} etc.) or telephone carriers (e.g., `` call
protect services offered by AT\&T~\cite{at&t}, Verizon~\cite{verizon} etc.) can
be helpful. However, these services typically rely on historical data such as user complaints or honeypot-generated
information~\cite{blacklist-paper,bharat-paper,c10}, and their
overall effectiveness tends to be low due to caller ID
spoofing~\cite{hiya-report}. 

\begin{comment}
Such spoofing has significantly increased recently (robocall
blocking company Hiya reported that 56.7\% scams reported by their users relied
on neighbor spoofing \cite{hiya-report}). Hence, solutions that rely only on
call source metadata in conjunction with blacklists cannot address the unwanted
robocall problem.
\end{comment}

\par 
A number of research papers have explored how caller ID spoofing can be detected
\cite{c3,c5,c6,c7}. However, the applicability of these approaches for protecting users from fraudulent calls has not been demonstrated.
One approach to better inform users about the sources of their calls is to limit or prevent caller ID spoofing, thus improving the efficacy of blacklists. The Federal Communications Commission (FCC) has mandated US telecom companies to start using SHAKEN/STIR by June 30,2021 \cite{fcc-june}. SHAKEN/STIR \cite{stir,shaken} enables the callee to verify the correctness of the caller ID. However,SHAKEN/STIR might not be as effective if most of the scammers are operating outside of US, or victims continue to fall
for scams that don't use caller ID spoofing \cite{doupe-article}.

Phoneypot \cite{phoneypot} demonstrated the feasibility of using a telephony
honeypot to augment abuse information. Other papers have explored vulnerabilities in
telephony systems, their exploitation and the motivation for telephony-based
attacks \cite{c7,c11}. In recent work that conducted a large scale user study
\cite{c13}, it was shown that a significant fraction of users fall victim to
telephone scams. \cite{ndss2020} evaluates the impact of user interface design elements on user decision-making for robocall blocking applications. In contrast, our work aims to handle incoming calls and provide a
solution to block unwanted calls from reaching recipients without user
intervention, even in the presence of caller ID spoofing.
\vspace{-1ex}
\section{System Design} \label{sec:sys_design}
\subsection{System Overview}
In this paper, we propose RobocallGuard, a virtual assistant (VA) based solution that can help defend against unsolicited phone calls. We developed a Smartphone app which can screen incoming calls without user interruption and intervention. The app hosts a VA, which works as a human secretary and receives incoming calls on behalf of the user. If the incoming call is from a whitelisted caller, the VA does not pick up the call and immediately notifies the user by ringing the phone. A whitelist can be defined by the user, and can include the user's contact list and other allowed caller IDs (such as a global whitelist which consists of public schools, hospitals etc.). On the other hand, if the call is from a blacklisted caller, the  VA blocks it and does not let the phone ring. However, if the caller ID belongs to neither a whitelist nor a blacklist, the VA picks up the call without ringing the phone and initiates a conversation with the caller to decide whether this call should be brought to the attention of the user. To make this decision, the VA presents the caller with a challenge which must be passed to reach the callee. The challenge can be thought of as an audio captcha which verifies the legitimacy of callers. To keep the caller experience natural, we experiment with a simple audio captcha, the name of the callee. Upon picking up the call, the VA asks the caller to state the name of the callee. If the caller says the correct name, the call is passed to the callee by ringing the phone. The transcript of the conversation between the VA and the caller is also shown on the phone screen to provide the callee with additional context. If the caller can not pass the above mentioned challenge, the VA blocks the call and notifies the user of the blocked call through a pop-up app notification. The VA also makes a decision if the call is from an unwanted human caller or a robocaller (discussed in detail in the later sections). Upon making this decision, the VA ends the call and stores the audio recording and transcript of the call for the user's convenience. Each audio recording and transcript is appropriately labelled (unwanted human caller or robocaller) by the VA. Since our proposed solution does not depend on the availability of a  blacklist of known robocallers, it can be effective even in the presence of caller ID spoofing.
\vspace{-1ex}
\subsection{Threat Model}
\subsubsection{In-scope Threats}
In this section we describe the scope of threats that our virtual assistant is designed to protect against.
\par \textbf{Mass robocalls:} Previous analysis shows that most of the robocall attacks that took place recently are mass calls. Attackers architect several campaigns such as tech support, IRS, free cruise and so on to reach a large number of phone users. Since the goal is to get a large coverage at minimal cost, attackers of such spam campaigns rarely target their victims individually. As a result, a simple audio challenge provided by the VA in the beginning of the call can filter out mass robocalls. Callers that cannot pass this challenge are not able to directly reach the callee. 
\par \textbf{Mass unwanted live calls from human:} Our defense mechanism not only protects against robocallers, but also from unwanted human callers such as telemarketers and debt collectors who repeatedly try to reach people at wrong phone numbers. Unless, the caller knows the name of the callee, the callee is not interrupted by the call and is only notified asynchronously via a message that includes call context.
\par \textbf{Spoofed Calls:} The use of caller ID spoofing has  increased significantly in the phone fraud eco-system. Neighbor spoofing is a common tool used by attackers these days. In our system, all incoming calls are picked up by the VA first, and audio analysis and speech recognition techniques are applied to prevent spoofed calls from interrupting the callee.
\par \textbf{AI equipped attack:} Attacks where the attacker is equipped with AI are not common in the phone fraud eco-system. However, with the availability of tools like Google Duplex \cite{duplex}, attackers can craft more sophisticated attacks where robocallers make a natural conversation with the other party and bypass our proposed defense mechanism. However, unless the attacker knows the name of the callee (which is not the case in mass robocalling campaigns), their call will not be passed to the user. The VA might mislabel an AI equipped robocaller as an \textit{unwanted} human caller, but will still be able to stop the unwanted call from reaching the user.

\subsubsection{Out of scope Threats}
Our VA does not protect against the following types of attacks. The attacks discussed below are currently not common, but may emerge in the future in an attempt to defeat intelligent phone call defense tools such as the one we propose in this paper.
\par \textbf{Targeted attack:} Our VA only protects against mass spam/scam calls. If the attackers obtain the callee’s name associated with a smart phone number through leaked private data, they can evade the VA by saying the correct name. Currently, such targeted attacks are rare, but the increase of leaked private information may pose such threat in the future. We discuss this in detail in Section \ref{sec:discussion}.
\par \textbf{Landline calls:} RobocallGuard only protects callees when they use a smartphone. Hence, malicious actors making landline calls using public directories are out of scope.
\subsubsection{Possible Evasions}
A robocaller might try to bypass our proposed defense mechanism in the following ways.
\par \textbf{Common name attacks:} A  common  name  attack  is  where  a  robocaller  plays  a prerecorded  message  of  carefully  chosen  common  names  to bypass the VA. Since we evaluated the VA where the challenge is the callee's name, such an attack could evade the VA if the user has an identical or similar name to any of the common names used  by  the  attacker. However, the challenge can be made more difficult for the attacker by requiring both first and last names.
\par \textbf{Master key attack:} Another possible evasion technique is crafting a keyword that can evade a large set of names. This keyword works as a master key that might fool the VA to accept the crafted keyword as the correct name.
\vspace{-2ex}
\subsection{Design Goals} \label{sec:goals}
In this section, we state the design goals that are required from a defense system designed to combat unwanted calls. Such a system needs to perform content analysis since relying only on caller ID is not sufficient to stop unwanted calls.
\begin{compactitem}
\item \textit{Add an extra layer of security between the caller and the recipient of the call.} This facilitates that all calls from {\em unknown} phone numbers (i.e., phone numbers not stored in the recipient's contact list) are passed through the VA, which filters out robocalls and unwanted human calls. The motivation behind this design goal to add a challenge to the caller before the callee's phone rings and  interrupts him/her. The challenge should be easy and natural enough for a legitimate caller to pass, but harder for mass robocallers to pass.
\item \textit{Provide additional information to the recipient about the content of the call}, so that they can make an informed decision to pick up the call. Such information should be extracted from audio received from the caller before the call is picked up by the callee.
\item \textit{Preserve user experience} in regards of latency and accuracy. The VA should not make the phone call experience too unnatural for the caller or callee.
\item \textit{Ensure privacy} when an incoming call is handled by the VA. In other words, the VA should run locally and the call is not transferred elsewhere. This ensures that the conversation between the caller and the VA is protected. 
\end{compactitem}
In our implementation of the VA we are able to achieve the first three goals. We also believe that ensuring privacy is feasible and we discuss it in detail in the later sections.

\begin{figure}[t]
\centering
\includegraphics[scale=0.3]{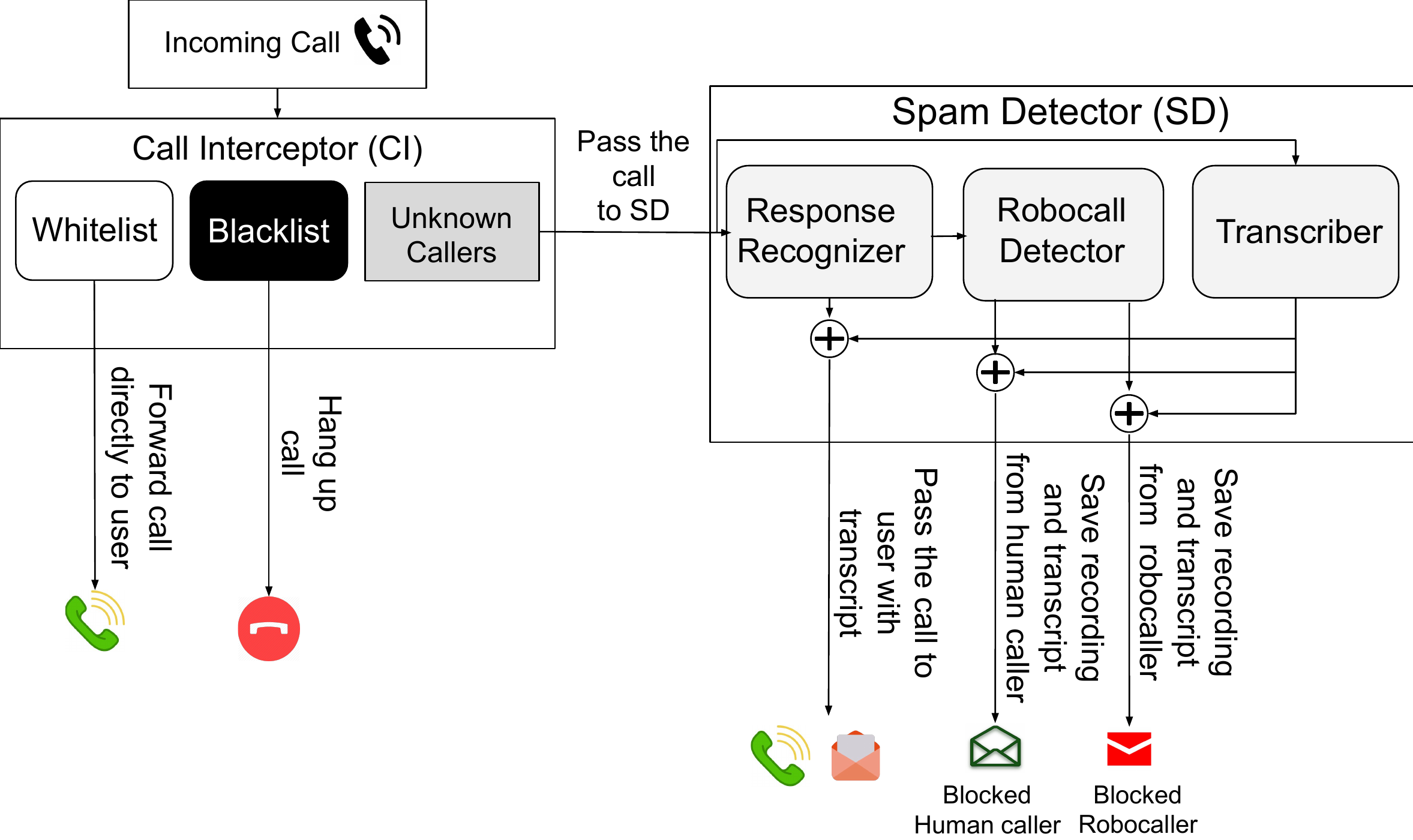}
\vspace{-10pt}
\caption{System Architecture}
\label{fig:system}
\vspace{-10pt}
\end{figure}

\subsection{User Workflow}\label{sec:workflow}
We consider a scenario in which smartphone users install RobocallGuard. All incoming calls are sent to the VA. Hence, the phone does not immediately ring and notify the callee of the incoming call (Fig: \ref{fig:system}). The VA makes a preliminary decision based on the caller ID of the incoming call. Whitelisted callers are immediately passed to the callee and blacklisted caller are blocked. The following steps are followed to handle calls from neither blacklisted or whitelisted callers.
\begin{enumerate}
\item The VA picks up the call, greets the caller and asks for the name of the person the caller is trying to reach. This enables the VA to check if the caller knows the callee, and if it does not, it is likely that this call is unwanted.

\item If the caller says the correct name, the VA detects it and passes the call immediately to the callee by notifying them of the incoming call. The correct name is set by the user when they install the app. The VA also provides the transcript of the conversation it just had with the caller to the user. While passing the call, the VA asks the name of the caller as well and provides it to the callee along with the transcript.

\item After a predetermined time t1 has elapsed from the initial greeting and the caller has not said the correct name, the VA asks for the name again and looks for caller interruption while playing this message. The intuition behind this is to differentiate robocallers from actual human callers. Current robocallers typically play a pre-recorded message and do not stop to make a conversation with the callee. Therefore, the callers who are not interrupted by the VA are labeled as robocallers. On the other hand, if the caller is silent during most of the time the VA is talking, we label them as potential unwanted human callers.
\item After t2 seconds have passed from the initial greeting and the caller has not said the correct name, the VA hangs up the call and saves the entire audio recording of the conversation it had with the caller. The VA also saves a transcript of the audio. The audio recording and transcript is labeled as robocaller or human caller according to step 4. Finally, the callee is notified of this blocked call and provided with the audio recording and transcripts. This allows the user to make a decision if they want to call back the caller or not.
\end{enumerate}

\subsection{VA Architecture}
In this section, we describe our system architecture which is independent of the implementation environment. The underlying architecture of our virtual assistant is designed to fulfill the goals mentioned in Section \ref{sec:goals}. To achieve our first goal of enhanced security, the modules, Call Interceptor and Spam Detector, act as a middle layer. They collect and analyze additional call information to enhance security against unwanted calls. The communication flow between the system components ensure that user experience is preserved. To ensure privacy, our envisioned system handles all calls locally. This could be achieved by implementing the VA as part of the default phone app, for example. In contrast, Robokiller~\cite{robokiller} routes all incoming calls to a central server, thus exposing possibly sensitive audio to a third-party. Figure \ref{fig:system} depicts the main components that make up our VA.
\par We envision our system to be embedded in the Phone app of a smartphone. The call screening feature provided by Google Pixel \cite{pixel} phones demonstrates the feasibility of embedding a VA with the phone app to intercept and examine voice from incoming phone calls. However, due to certain OS enforced restrictions, we implement a proof-of-concept prototype instead of embedding the VA with the Phone app. Such limitations and the choices we make to overcome them are discussed in Section \ref{sec:implementation}. In the following, we describe each component of the system architecture.

\subsubsection{Call Interceptor}
All incoming calls are passed to the Call Interceptor(CI) module. The main function of the CI module is interception of a call to acquire the incoming audio stream, and injection of recorded voice messages by the VA into the outgoing audio stream. The CI makes an initial decision based on the caller ID. All calls from whitelisted phone numbers are passed to the user (callee in this case) without further processing. A user has total control of the whitelist on his or her phone and can decide phone numbers from which calls should be passed to them directly without intervention from the VA. All calls from blacklisted phone numbers are dropped and stopped from reaching the user. The blacklist is predefined as well; however, it is designed to be dynamic to include newly appearing malicious phone numbers \cite{blacklist-paper}. Phone numbers which are not present in the whitelist or blacklist are labeled as "Unknown Callers". Audio stream from all unknown callers are passed to the Spam Detector module for further analysis.
\subsubsection{Spam Detector}
This module analyzes the audio coming from an unknown caller to make a decision about the nature of the incoming call.
\vspace{-2ex}

\paragraph {Response Recognizer}\label{sec:name_recognizer}
\par Incoming audio stream from calls originating from unknown callers are passed from the Call Interceptor to the Response Recognizer(RR) module. The RR module decides whether to pass the incoming call to the callee. If the call is considered unwanted, it is handled by the VA and not passed to the user. However, since the notion of an unwanted call is different for each user, it is difficult to define an unwanted call. Hence, we take a conservative approach: if the caller knows the name of the callee, we label that call as \textit{wanted}; conversely, when the caller does not know the name of the callee, that call is labelled as \textit{unwanted}. The intuition behind this approach is that phone calls from a person who knows the callee are less likely to be unwanted.
\par The user is allowed to set the name(s) that should be accepted as correct by the VA. We refer to the name(s) set by the user during the installation of the app as \textit{correct name(s)}. Users may set multiple \textit{correct names} as a valid recipient of phone calls coming to their device. After the installation, during a future phone call, the caller is asked who they are trying to reach. After the call has been picked up, we set a limit of 35 seconds (value of time limit $t_2$) to allow the caller to say a \textit{correct name(s)}. The value of $t_2$ is set empirically, keeping in mind that the VA speaks for 15 seconds during the 35 second time limit, hence the the remaining time period should be enough for the caller to provide meaningful context. Moreover, $t_2$ can be tuned to match user experience and context details. If the caller says any of the \textit{correct names} at any point during this 35 second period, the RR module recognizes it and passes the call to the user along with the transcript of the conversation between the VA and the caller. However, if the caller does not say any of the \textit{correct names}, the call is deemed as unwanted and not passed to the user.
\par The backbone of the RR module is a keyword spotting algorithm which can detect the right keyword. In our scenario, the correct name(s) of the callee is the keyword. There has been a lot of research on keyword spotting algorithms which are used in many commercially available products such as Amazon Alexa, Okay Google in Google products etc. Hence, we explored existing systems that can effectively detect a keyword. However, for such a system to be usable in our VA, high accuracy with limited training examples, an open source toolkit and  a light enough system to run on a mobile device is required.

\par Since the users set the \textit{correct name} by making audio recordings of them pronouncing the names, it is not feasible to collect a large number of audio samples from the users. In other words, the keyword spotting algorithm will have access to only a few recordings of each name. However, the system should have a high true positive rate, and a low false negative rate. Snowboy \cite{snowboy}, CMU Pocketsphinx \cite{cmu}, Honk \cite{honk} are all open source keyword spotting toolkits that are light enough to run on a mobile device. Based on  our experiments, we found CMU Pocketsphinx has lower accuracy than Snowboy, when trained with names. Honk requires a larger number of audio samples to train a keyword. On the contrary, Snowboy requires only 3 audio recordings to train a keyword. Snowboy also supports multiple keyword models, thus multiple names can be set as keywords. Hence, we chose Snowboy to recognize the name. Because Snowboy does not connect to the Internet, it can ensure privacy, which is one of our design goals. We treat Snowboy as a blackbox, which when provided with 3 audio samples, creates a model to detect the keyword. We embedded the downloaded trained model with the VA to recognize the \textit{correct name(s)}.

\vspace{-2ex}
\paragraph{Robocall Detector}\label{sec:robocall_detector}
The objective of this module is to determine whether the caller is an actual human or a robocaller. When a caller cannot say the correct name, their call is handled by the Robocall Detector(RD) module.

\par As mentioned in the workflow Section \ref{sec:workflow}, the RD is activated after $t_1$ seconds. We set $t_1$ to 20 seconds as we want to give the caller enough time to say the correct name. The analysis of the data we collected from our user study shows that the initial 20 seconds is a reasonable time for the callers to say the correct name even if they have to repeat the name. After 20 seconds have passed from the initial greeting and the caller has not said the correct name, the VA plays an audio message to interrupt the caller. Let the duration of this audio played by the VA be $t_3$ seconds (we set $t_3$ to 5 seconds in our experiment). During these $t_3$ seconds, the RD module checks if there is silence from the caller's side. We use Voice Activity Detection (VAD) \cite{vad} to determine if the audio coming from the caller's side contains voice or silence. If the caller is silent for at least $t_3$/2 seconds while the VA is playing the audio message, we label the caller as an actual person. On the contrary, if the caller is silent for less than $t_3$/2 seconds, it is labelled as a robocaller. Determination of the type of the caller (human or robocaller) provides additional information to the user about the call. Upon determining the type of the caller, the VA allows the caller 10 more seconds as a margin of error to say the correct name before hanging up the call. The audio recording and transcripts of the entire conversation with the caller is saved locally at the device for the user to preview later. The associated label (human or robocaller) is used to determine in which folder the audio and transcript is stored. 
%\vspace{-3ex}
\paragraph{Transcriber}
The transcriber component transcribes the entire conversation between the VA and the caller to provide additional context to the user. The VA stores the audio recording and the transcription of that audio recording locally; and notifies the user of these two files after it has handled the incoming call. This helps the user to access the content of the call without picking up and engaging in the phone call. Calls that are passed to the user by the VA after deeming them as ``wanted", are also provided with a transcript of the conversation between the caller and the VA that took place before the call was forwarded to the user. When the user is notified of a  such a call by ringing the phone, the transcript is shown on the screen for additional context. Calls which are deemed ``unwanted", both from human callers and robocallers, are not passed to the user and hung up by the VA.
\par The VA does not engage in a conversation with callers that are whitelisted and passes the calls directly to the user. Therefore, transcripts are not provided for such calls. All other callers are greeted by the VA and hence transcript is made available to the user to understand the content of the calls.
\par There are many software libraries and APIs available for transcription. We have chosen Google Cloud Speech  API \cite{google-speech}  because it has a very high accuracy and transcription can be performed from a mobile device, unlike Kaldi \cite{kaldi} and Mozilla deep speech \cite{deep-speech}. Ideally, the transcription should be conducted locally in the device and no server should be involved. Android provides the means to perform transcription locally through \textit{RecognizerIntent} and \textit{SpeechRecognizer} class. However it imposes the restriction of the input channel. \textit{RecognizerIntent} and \textit{SpeechRecognizer} class always uses audio from the microphone as an input for transcription. As discussed in Section IV, audio from the caller in our implementation comes though the VoIP channel instead of the microphone. Therefore, in our proof-of-concept prototype, we do not perform transcription locally. Instead, we send the stored audio recording of the conversation between the VA and the caller to Google Cloud and a corresponding transcript is returned. However, when RobocallGuard is deployed in real life, transcription can be done locally in the device, thus privacy can be ensured.
\vspace{-1ex}

\section{Implementation} \label{sec:implementation}
In this section, we discuss some important details of our implementation of the VA. We implemented a prototype of our app using Java on Android. We envision our VA to be embedded with the Phone app where the VA handles all incoming calls locally without having to stream audio to an external server. The recent release of call screening feature for Android phones further supports the feasibility of such a system. However current Android system restrictions do not allow embedding a system that can inject voice messages in the outgoing audio stream without OS modifications. Hence, in the following, we describe and explain the choices we made in the implementation of our proof-of-concept prototype.

\subsection{Motivation behind VoIP} The workflow of the VA starts with an incoming call being passed to the CI module. As discussed earlier, the CI module captures the audio stream from caller side and takes full control of the stream so that the system is capable of analyzing the voice data, as well as locally recording the caller's audio. Moreover, the CI also injects audio into the phone call to communicate with the caller on behalf of call recipient, while the recipient has no awareness of the incoming call during the time the VA is interacting with the caller. To perform the first task, we could implement a customized phone call app by making modifications to phone call service codebase provided by Android (i.e. Implementing customized Android.telecom.InCallService). However, it is strictly constrained for a common developer to satisfy all the requirements for injecting audio to an ongoing phone call. For the sake of security and privacy, Android does not allow injecting sound files in the conversation during a phone call \cite{android}, which means that no such API is provided by the Android system that could pre-process or replace the microphone as an input audio stream during a phone call.
\par Taking all these into account, we decided to implement a VoIP(Voice over Internet Protocol) application to conduct our user study experiment. With an VoIP phone call application, we are able to get full access to voice streams on both sides of a phone call. Furthermore, it is possible to inject audio at any appropriate moment in the conversation during a phone call with a VoIP app.
\vspace{-1ex}
\begin{figure*}[t]
\centering
\includegraphics[scale=0.37]{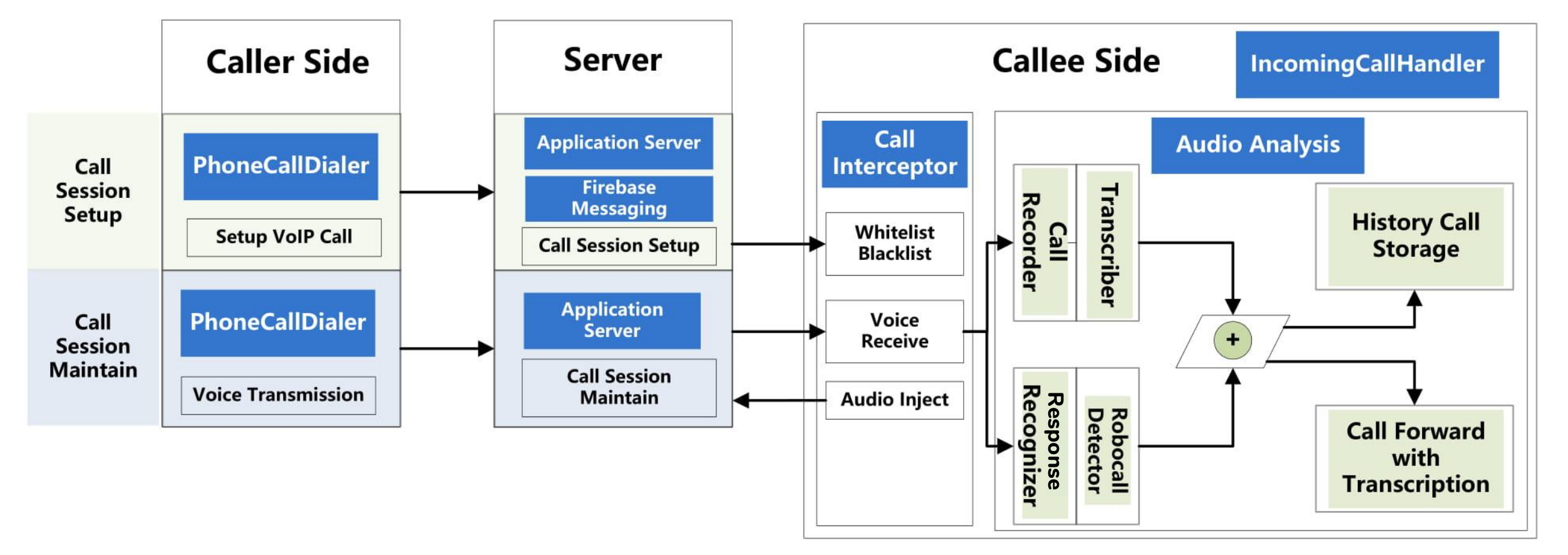}
\vspace{-10pt}
\caption{VoIP Architecture}
\label{fig:voip}
\vspace{-10pt}
\end{figure*}
\subsection{VoIP application architecture}
\par Our VoIP application consists of two major parts, namely client-side and server-side systems. For the sake of convenience, we implemented the system so the two parts work over TCP network connections by using a simplified and customized session initiation and keep protocol. Figure \ref{fig:voip} demonstrates the architecture of our VoIP application.
\par On the server side, an application server (Server) working along with Google Firebase Cloud Messaging Platform (FCM) \cite{firebase}, handles call session setup and maintaintenance. On the client-side, we split the core of our call implementation into two parts, one called ``PhoneCallDialer (PCD)" and the other called ``IncomingCallHandler (ICH)". PCD is mainly a dial service, which provides a dialing panel for the user to make VoIP phone calls. At call session setup stage, PCD initiates a connection request and builds a call session between caller and callee with the help of the Server. At call session maintain stage, PCD streams and transmits caller's voice to callee through the channel maintained by Server. 
\par ICH consists of two functional modules, which implement callee-side handling of incoming calls. The CI module automatically answers an incoming call without the callee's awareness during call session setup stage and makes the initial decision to forward calls from whitelisted numbers or deny calls from blacklisted numbers. For unknown callers, the call is passed to the Spam Detector (SD) module and corresponding pre-recorded audio tracks are injected into the call to allow the VA to communicate with the caller.

We implemented and used the system described above to conduct a user study and evaluate the efficacy of the VA. The main purpose of developing a VoIP prototype is to assess the usability of the VA and its effectiveness in detecting robocalls and other unwanted calls. Although we developed a VoIP prototype, it is possible to embed our VA in the Phone app.
\vspace{-1ex}
\section{VA Evaluation}\label{sec:results}
In this section, we report the results of experiments we conducted to measure the accuracy of decisions made by our VA for incoming calls, and discuss a user study that was conducted to evaluate the usability of our prototype.

\subsection{Usability Study}\label{sec:user_study}
Our VA is designed to provide the convenience of a human assistant while detecting unwanted calls. It also provides context for calls, which helps the callee decide if a call needs his/her attention. To explore the usability of such a system, we conducted an Institutional Review Board (IRB) approved user study. In the following, we first describe the study setup, its participants and then discuss the results.
\vspace{-0.5ex}
\subsubsection{Study Setup}
Our study participants consists of 21 users who were sampled from a population of college students. Most of the participants can be described as tech-savvy. All participants were required to be fluent in English and be familiar with using smartphones. Each experiment was conducted with two Android devices, a Samsung Galaxy S9 plus and a Samsung Galaxy Tab A, running Android 8.0 and 7.0 respectively. Both of the devices had RobocallGuard app installed. During the user study, all phone calls were made using the Tab and received using the smartphone. The setup of the experiment is as follows. We briefed the participants about the experiment process and explained the purpose of the VA. We asked the users to perform a list of tasks: making calls, receiving calls, checking the contents of blocked calls and answering multiple choice questions regarding their experience of using RobocallGuard. Participants were assigned the role of a caller and a callee one at a time and were asked to make/receive a call. When participants were assigned the role of a caller, they made 4 calls. Such a call took at most one minute. When participants were assigned the role of a callee, they received 5 calls. After making/receiving each call, the participants had to answer multiple choice questions regarding their experience. At the end of the user study, each user was asked 6 generic questions about their overall experience with the app.
%\vspace{-3ex}
\subsubsection{User Actions}
In this section, we describe each task the participants performed during the user study in detail. Each experiment involved a pair of users (user A and user B), one caller and one callee, performing the tasks. Once user A has completed all the tasks assigned to the caller, they are assigned the role of a callee and vice versa. The experiment starts with user A acting as the caller and making calls to the callee, user B.
\par We performed two sub-experiments within each experiment. During sub-experiment 1, we provided the caller with the appropiate response to the challenge i.e. the correct name. Hence, the caller should be able to reach the callee. Conversely, in sub-experiment 2, the caller is either given an incorrect name or no name at all. Therefore, the VA would not allow them to reach the callee. The participants had no idea about what the correct name was. Furthermore, there are two scenarios in each sub-experiment; one where the caller is provided with a script to read from when making a conversation with the VA, and one where the caller is given a topic to talk about, instead of a script, while interacting with the virtual assistant (e.g calling a friend to make movie plans.) When a user is assigned the role of a callee, with each forwarded call they are given the choice to either pick up or decline the call. They are advised to use the caller-VA interaction transcript provided by the app to make this decision. Once they pick up the call they can start a normal conversation with the caller. During our user study, we preset the correct name to be \emph{Taylor} instead of having each user set a name. We make this choice because the purpose of the user study is to get insights about call experience in the presence of a VA, rather than testing the accuracy of the keyword spotting algorithm. We ask the callee to impersonate Taylor when making the decision to answer an incoming call. Furthermore, the callee has no advance knowledge of the content of the call. Following is the detailed description of the 4 calls made by the caller during the experiment.

\textbf{Sub-experiment 1:} In this case, the correct name ``Taylor" is provided to the caller and the callee is asked to impersonate Taylor.

\textit{Scenario A:} When asked the name of the callee, the caller is instructed to read the following script, ``Hello, can you please forward my call to Taylor?". In this scenario the VA allows the caller to reach the callee.

\textit{Scenario B:} In this scenario, the caller is not given a script. They are instructed to make a call to their friend Taylor to make movie plans. They are advised to include the name Taylor in their conversation.

While forwarding the call to the callee, the VA asks the caller to state their name. We advised our participants to say a fake name to protect their privacy.

\textbf{Sub-experiment 2:} In this case, an incorrect or no name is provided to the caller and similar to sub-experiment 1, the callee is asked to impersonate Taylor, the correct name being set as Taylor.

\textit{Scenario A:} In this scenario the caller is asked to read the following script, ``Hello, can you please forward my call to Robert? We met at a seminar today." It should be noted that Robert is the incorrect name here, hence the VA blocks this call.

\textit{Scenario B:} In this scenario, the caller is not given any script or name. They are instructed to make a call to an office trying to sell a computer. Since the caller does not say the correct name during this call, the VA blocks this call as well.

As a result the callee gets a notification of the blocked call along with the transcript of the call. After making each call, the caller is required to answer survey questions that focus on the ease of interacting with the VA, the quality of the transition from the VA to the callee and the delay experienced before the callee responded. As a callee, a participant received the aforementioned 4 calls. The VA passes the first two calls to the callee since the correct name was said by the caller. On the other hand, the VA blocked the last two calls, notified the callee of the blocked calls and provided her with the transcript and audio recording of the conversation. In addition to receiving the 4 calls made by the caller, the participant received a robocall made by us. The VA blocks this call, labels it as a robocall and notifies the user. After each call (both blocked and passed), the callee is required to answer a number of survey questions regarding the usefulness of the transcript, the interaction with the caller, and the reason behind their decision to pick up or not pick up the call.
 
 \begin{figure*}[t]
\centering
\begin{subfigure}{.5\textwidth}
  \centering
  \includegraphics[width=2.5 in]{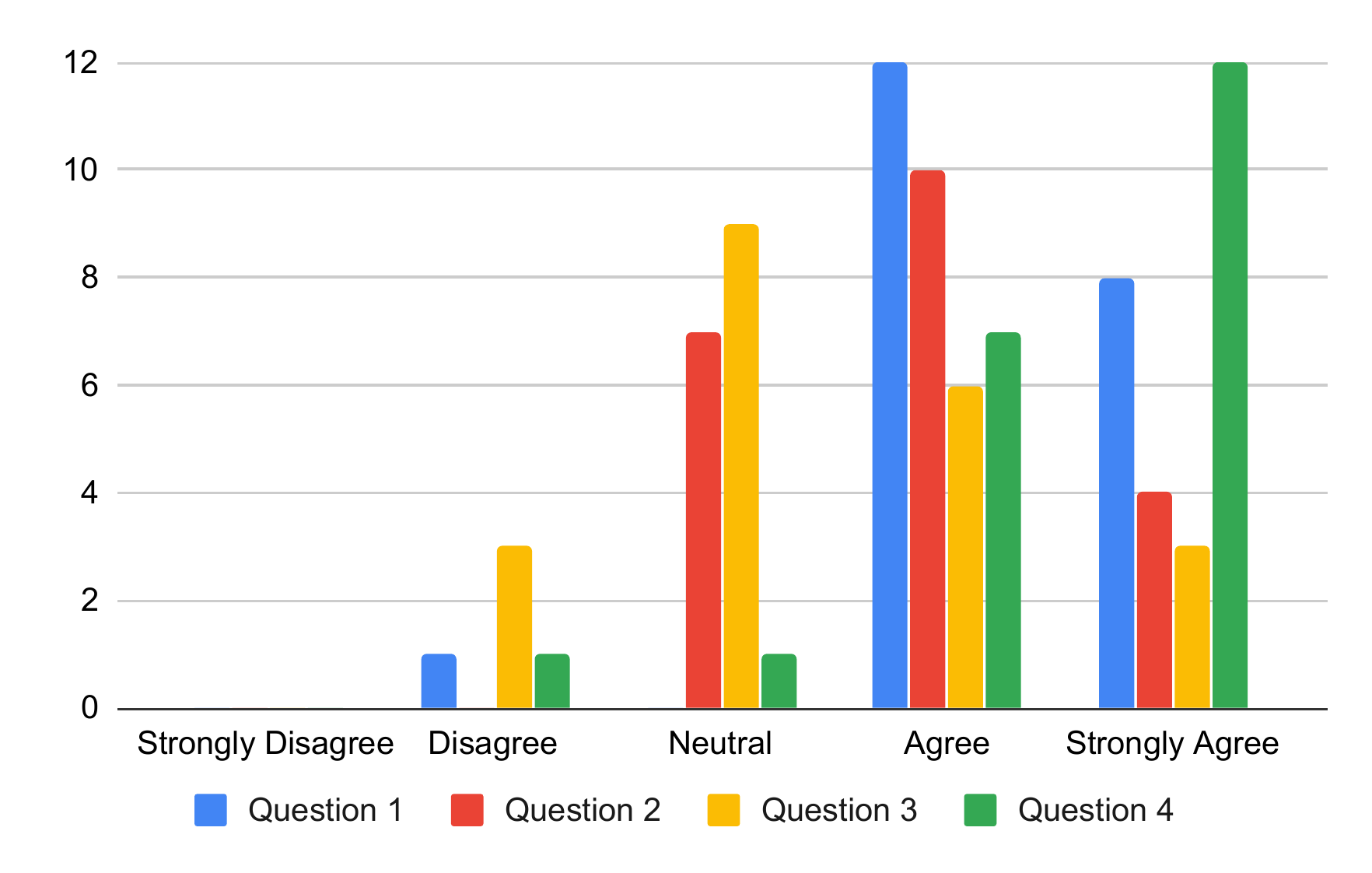}
  \caption{Caller and Callee Response}
  \label{fig:caller_callee}
\end{subfigure}%
\begin{subfigure}{.5\textwidth}
  \centering
  \includegraphics[width=2.5in]{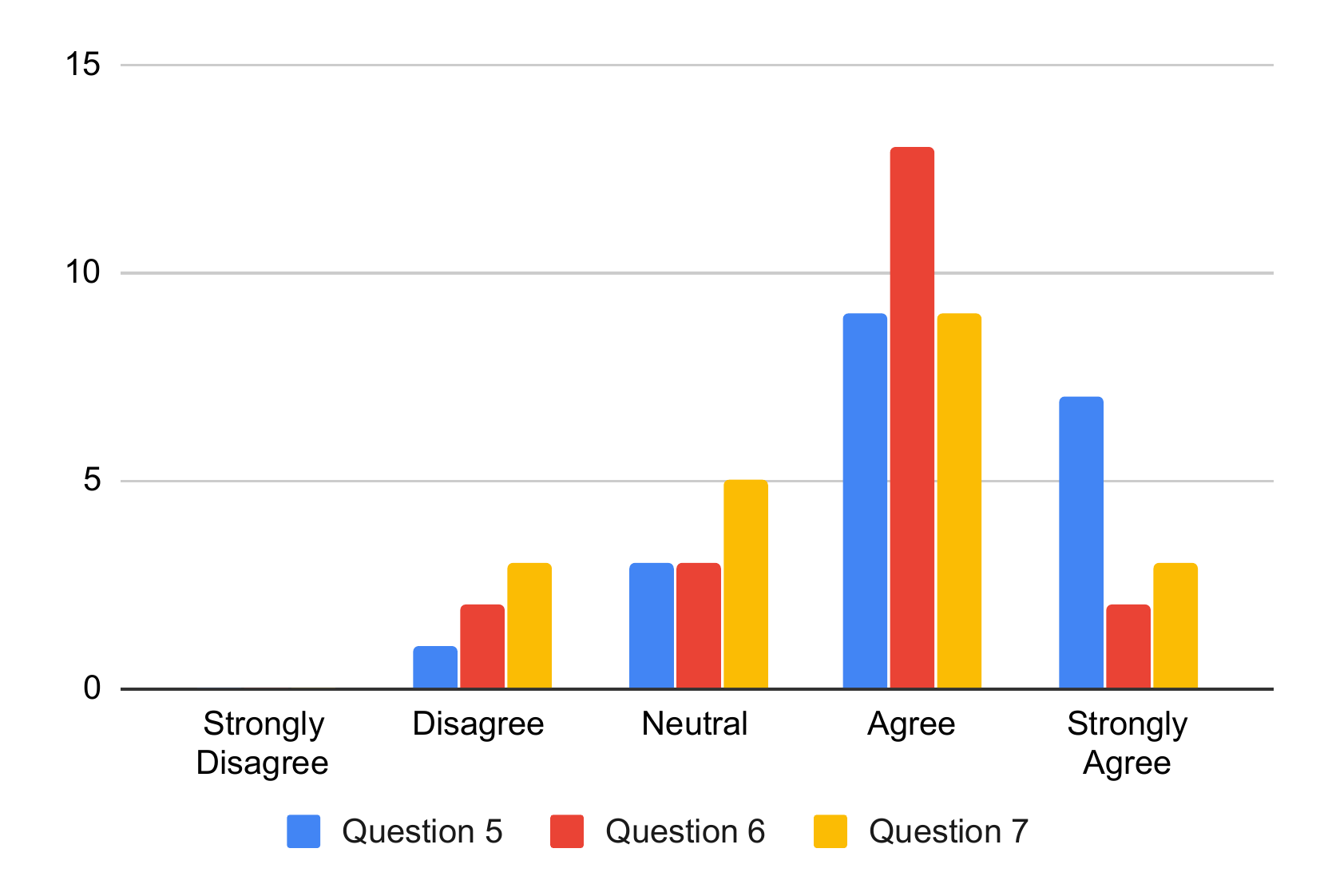}
  \caption{User Generic Response}
  \label{fig:user_generic}
\end{subfigure}
\caption{User Results}
\label{fig:user_results}
\vspace{-10pt}
\end{figure*}

\subsubsection{User Study Results} \label{sec:user_study_results}
In this section we present the results obtained from the user responses to the survey questions. We present user responses to the following seven questions in Figure: \ref{fig:user_results}. Questions 1-2, 3-4, 5-7 investigate caller, callee and overall user experience respectively.
\begin{itemize}
    \item Question 1: It was easy to interact with the VA.
    \item Question 2: The delay you experienced before the other person responded to the call is acceptable.
    \item Question 3: The transcript was able to provide sufficient information to infer the topic of the incoming calls.
    \item Question 4: The transcript was able to provide sufficient information about the content of the blocked calls.
    \item Question 5: I found the app beneficial to me as it provides prior knowledge about the incoming calls.
    \item Question 6: I think I would like to use an app equipped with a VA frequently.
    \item Question 7: I felt comfortable with the VA intervening in the phone calls.
\end{itemize}

\textbf{Caller Experience:} Most of the users who acted as the caller reported that it was easy to interact with the VA. Figure \ref{fig:caller_callee} shows the distribution of the responses to a question about the ease of interaction with the VA when the caller said the correct name. When the caller was given an incorrect or no name (in case of sub-experiment 2), 4 out of 21 users disagreed that it was easy to interact with the VA. This is understandable since the user had no knowledge of what the correct name is, they became frustrated when the VA did not pass their call to the callee, which is what may happen when a telemarketer calls. Figure \ref{fig:caller_callee} also shows the user responses regarding the delay experienced by the caller before the callee could be reached. Most users agreed or strongly agreed that the delay was acceptable. Moreover, all but one caller reported that they were satisfied with the transition from the VA to the callee. Hence, it can be concluded that a call experience for callers was not degraded as a result of having the VA acting as an intermediary.

\textbf{Callee Experience:} While acting as the callee, when a call is passed to the user, they may either answer or decline the call. Most of the users who picked up the call said that they made this decision because the transcript suggested this to be a non-spam call. Only 4 users reported that they answered the call because they pick up all incoming phone calls. All but one callee said that the interaction with the caller felt normal/natural after they picked up the call. Figure \ref{fig:caller_callee} shows how the callees felt about the transcript of the incoming calls. Only 3 users reported that the transcript was not able to provide with sufficient information about the incoming calls. The reason behind this is that some callers only said the correct name ``Taylor" and nothing else after starting a conversation with the VA. Thus, the transcript consists of only the correct name and no other information. Since it is not possible to control what the caller says, the quality of the transcript cannot be guaranteed in all cases. However, since the caller is also required to say their own name before a call is forwarded, it provides further information about the call to the callee. The current version of our prototype does not ask the caller to state the purpose of the call. We plan to add this feature in our future version, which we expect will augment the content of the transcript. Figure \ref{fig:caller_callee} also shows callee responses about the transcript of the blocked calls. All callees reported that they were notified in a timely manner of a new blocked call.

\textbf{Overall Experience:} At the end of the user study, each user was asked 6 generic questions. According to Figure \ref{fig:user_generic}, most users mentioned that they found the app beneficial to them as it provides prior knowledge about incoming calls. Moreover, most users said that they would like to use an app equipped with a virtual assistant frequently. In addition to these questions, all but two users said that the app was easy to use. Only 3 users reported that they needed to learn a lot of things before they could get going with this app and only two users reported that they could not easily navigate through the app. Moreover, all but 3 users reported that they were comfortable with the VA intervening in their phone calls before the calls are forwarded to the callee. Based on these results, we can conclude that the the VA does not negatively impact the overall call experience and lack of usability will likely not be the reason for impeding its adoption. Since we recruited 21 users for this study, the accuracy/quality of our conclusion from this study lies between 80\% to 95\% (between 10 and 20 users) according to \cite{user-study}.

%%%%%%%%%%%%%%%%%%%%%%%%%%
%%%%%%%%%%%%%%%%%%%%%%%%%%
\vspace{-1ex}
\subsection{VA Performance}\label{sec:exp_res}
In this section, we evaluate the effectiveness of the VA based on its ability to forward legitimate calls and block and label unwanted calls. For each incoming call, the VA has to make a decision whether to forward the call to the user or not. Besides, it needs to label each unwanted caller as human or robocaller.
\subsubsection{Interaction with Human Callers}
We first measure the effectiveness of the VA when the caller is a human. We categorize all human callers into two catagories: legitimate callers and unwanted callers. We define a caller to be legitimate if they know the correct name of the callee.  On the contrary when the caller fails to pass the challenge, we define them as unwanted callers. To measure the VA's effectiveness when callers are human, we used the data collected during the user study. We have discussed usability insights in our user study results in Section \ref{sec:user_study_results}; here we present statistics on the VA's accuracy.
\par \textbf{Legitimate callers:} As discussed earlier, during the user study, we performed four experiments with each of the 21 callers. In the first two experiments the callers were given the correct name of the callee. Our analysis shows that when callers passed the challenge, the VA detected all of them as legitimate callers and forwarded their calls to the callee. In only two cases, the callers had to repeat the correct name, and after the second time the callers had said the correct name, the VA forwarded the calls to the callee. This was mostly due to snowboy, and advancements in keyword spotting algorithms will further reduce such false negatives.
\par \textbf{Unwanted Callers:} During the user study we conducted, in the third experiment the callers were given an incorrect name and in the last one the callers were not given any name at all. Therefore, in these cases, the callers in our user study played the role of an unwanted caller in a way similar to a telemarketer, and made calls to the callee. The VA picked up the calls and was able to prevent every incoming unwanted call from directly reaching the callee. 
\subsubsection{Interaction with Robocallers}
In this experiment we measure the fraction of robocalls stopped and correctly labelled by the VA. We use a dataset of phone call records collected at a large phone honeypot provided by a commercial robocall blocking company. This dataset contains 8081 calls (which had an average duration of 32.3 seconds) coming into a telephony honeypot during April 23, 2018 and May 6, 2018 \footnote{Although this dataset is not recent, robocaller behavior has not significantly changed.}. It records the source phone number that made the call, the time of the call, the audio recording of the call and the transcript of the audio recording. Since it is a telephony honeypot, it contains some misdialed calls along with robocalls.
\par \textbf{Robocall Detection:}
To filter out misdialed calls, we use the approach previously proposed by Pandit et al. in \cite{blacklist-paper} to extract important topics from the transcripts of calls in our robocall dataset. Using LSI topic modeling, we extract 60 topics from our corpus of transcripts where each topic represents a spam campaign. We construct a similarity matrix by computing the cosine similarity between each transcript. We then convert the similarity matrix into a distance matrix by inverting the elements of the similarity matrix. We performed DBSCAN clustering on the distance matrix. DBSCAN is one of the most common clustering algorithms which given a set of points, groups together points that are closely packed together, marking as outliers points that lie alone in low-density regions. At the end of this, 79 clusters were created where each cluster represents groups of highly similar transcripts of robocalls. Since, the honeypot by nature contains unwanted calls, the clustering technique acts as a sieve that filters out all non-spam calls as outliers. Moreover, it allowed us to take one representative robocall from each cluster and use the audio recording to make a call to the VA. Upon making the calls, the VA correctly detected 100\% of all robocalls as unwanted and stopped them from ringing the user's phone. Theoretically, a false negative occurs when the VA forwards an unwanted call i.e. recognizes an incorrect response to the challenge as a correct one. A false positive occurs when the VA blocks a wanted call i.e. fails to detect the correct response to the challenge. With stricter challenges and fast advances in AI, we expect the correctness of the VA to increase.
\par \textbf{Robocall Labeling:}
 Once unwanted calls are detected, our VA determines if the caller sounds like a human or a robocaller. From the previous experiment, we noticed that the robocalls with a duration of less than 20 seconds were being inaccurately labeled as unwanted human callers. The reason lies in the design of the VA. The VA interrupts a caller at the 20th second and determines if they were interrupted or not by looking for voice activity from the caller side while the VA is speaking. The robocalls that are less than 20 seconds are found to be silent when the VA interrupts and are mislabeled as human. To solve this problem, we analyzed the contents of the short length robocalls. Since the robocallers are trying to financially profit from their victims, the content of the short robocalls must serve a purpose. From analysis of the robocall recordings in the honeypot, we discovered that 86\% of the short robocallers ask the callee to press or enter a digit in the phone keypad. Hence we take a further step to identify the short robocalls. If the transcript of a call contains the keywords "press" or "enter", we label it as a robocall. It is unlikely that a legitimate human caller will say these words while interacting with the VA. With this added step, our results show that the VA is able label 97.8\% of all robocalls correctly.
% \vspace{-1.25ex}
\subsection{Comparison with Call Blocking Apps}
There are several commercial applications available in the app stores that aim at blocking robocalls. Most of these apps (such as Youmail, Hiya, Nomorobo, etc.) rely on phone blacklists. These blacklists are generated from user complaints of unwanted calls (e.g., FTC reports) and other data collected by the app vendors.  Thus, these apps block an incoming call by looking at its caller ID and not the content of the call. As discussed earlier, spoofed calls, which are common, cannot be blocked with this approach. Currently, only Robokiller claims to perform call content analysis, in addition to using phone blacklists, to block unwanted calls. We performed a small scale experiment to compare Robokiller with RobocallGuard. To conduct this experiment, we installed Robokiller on a Samsung Galaxy S9 Plus Android device. We then used a Twilio \cite{twilio} phone number to make  10 robocalls to the device where Robokiller is installed. We chose a random sample of 10 robocall messages. Since the phone number we used is a Twilio verified phone number, it is not a blacklisted phone number. Therefore, a defense system that relies only on phone blacklists will not be able to block these robocalls. Only a system that performs audio content analysis will be able to detect these robocalls. However, we noticed that Robokiller was not able to block any of the robocalls made from the Twilio phone number and let the calls pass to the user without any spam call warning. In contrast, RobocallGuard was able to block all of these 10 robocalls. Therefore, it can be inferred that the available Robokiller app seems to rely mostly on phone blacklists rather than call content analysis. To explore this further, we downloaded FTC user complaint reports from August 3-5, 2019 and extracted 10 phone numbers that had most complaints. We then spoofed each of these 10 caller IDs using SpoofCard \cite{spoofcard} to make phone calls to Robokiller. Since we conducted this experiment on September 10, it can be expected that the top 10 callers from a month old FTC dataset will be in blacklists used by commercial apps. In addition, we performed reverse look up on each of these 10 phone numbers and found 7 of them to be labeled as ``scam or fraud". Upon making the spoofed phone calls, we found that Robokiller was able to block calls from 9 of the 10 caller IDs. This shows that Robokiller is able to block most of the incoming calls from blacklisted caller IDs.

In our next experiment, we downloaded FTC user complaint reports from September 9, 2019 and extracted the 10 phone numbers that were least complained about. We spoofed each of these 10 caller IDs to make phone calls to Robokiller. Since this dataset contained complaints from the previous day, it can be assumed that these phone numbers are not present in blacklists. Upon making the spoofed calls, we found out that Robokiller let calls from 8 of the 10 caller IDs pass. Although this is a small scale experiment, it provides evidence that Robokiller appears to rely more on phone blacklists and not the content of the call. Hence, unlike our proposed VA, robocalls with spoofed or previously unseen caller IDs can evade Robokiller.
\vspace{-1ex}
\section{Discussion}\label{sec:discussion}
RobocallGuard is the first version of our system to mainly test the usability of a VA and its effectiveness in stopping current robocalls. The survey responses from the user study we conducted show that both callers and callees are comfortable with the change in the call experience due to the VA and found the VA beneficial. Moreover, \cite{ndss2020} has shown that users need a app which handles spam calls without making the phone inoperable. The fact that RobocallGuard filters out spam call without user intervention and without making the phone inoperable adds desired convenience to the users. Much of the current voice abuse over telephony is perpetrated by mass robocallers who indiscriminately call a large number of potential victims. They use techniques like neighbor spoofing to increase the likelihood that their calls are picked up. Our results show that RobocallGuard can be effective against such mass robocallers.

We understand that the callee's name might seem like a simple challenge, which could be evaded by bad actors by obtaining names associated with phone numbers from leaked data. However, most of the robocallers currently make cheap mass robocalls. Hence, adding a simple challenge like the callee's name, adds cost to the malicious actors and works effectively to stop current robocalls. In our future work, we plan to explore stricter challenges, such as, interrupt and make conversation with the robocallers, ask further questions that are easy for a legitimate caller to answer but difficult for a robocaller. Such challenges would be difficult to break for a more sophisticated robocaller without AI capabilities. 
The VA based defense proposed by us has a few limitations. It only works when the callee has a smartphone. The user study and performance evaluation experiments were conducted with a specific name set as the correct name. Since evaluating the correctness of keyword spotting algorithms is out of our scope, we did not conduct experiments with a broader range of names. Also, our user study is limited to tech savvy university students. We hope the results of the study would be applicable to the broader population.

An important feature of our VA is that it can also stop unwanted live calls that come from human callers such as telemarketers. Although we ensured that the VA interacts with live callers in our user study, the interaction occurred in a controlled environment. In the future, we plan to conduct experiments in which the VA handles unwanted calls that come from live sources \cite{lenny}. 
\section{Conclusion} \label{sec:conclusion}
In this paper, we proposed RobocallGuard, a virtual assistant (VA) system that aims to automatically detect and block robocalls before they reach the targeted user. We developed an Android prototype app and conducted a user study, and showed that the VA can effectively block unwanted calls without disrupting the caller or callee experience. We also showed that our VA is able to correctly label 97.8\% of robocalls without negatively impacting legitimate calls. 
\vspace{-1ex}

%\section*{Acknowledgment}



%\bibliographystyle{IEEEtranS}

\end{document}